\documentstyle[12pt]{article}

\def\frac#1#2{{\textstyle {#1 \over #2}}}

\def\Eq{\begin{equation}}   \def\Endl#1{\label{#1} \end{equation}}
\def\Eqa{\begin{eqnarray}}  \def\Endla#1{\label{#1} \end{eqnarray}}
\def\Enda{\end{eqnarray}}
\def\End{\end{equation}}

\def\O{ {\cal O} }

\def\&{and}

\def \PS {p\!\!\!/}
\def\ap#1#2#3{           {\it Ann. Phys. (NY) }{\bf #1}, #2 (19#3)}

\def\np#1#2#3{           {\it Nucl. Phys. }{\bf #1}, #2 (19#3)}
\def\pl#1#2#3{           {\it Phys. Lett. }{\bf #1}, #2 (19#3)}
\def\pr#1#2#3{           {\it Phys. Rev. }{\bf #1}, #2 (19#3)}

\begin{document}
\begin{titlepage}

\begin{center}
June, 1993      \hfill       HUTP-93-A019\\
\vskip .5 in
{\large \bf Nonperturbative Decoupling and Effective Field Theory}
\vskip .3 in
{
  {\bf Stephen D.H. Hsu}\footnote{Junior Fellow, Harvard Society of
     Fellows. Email: \tt Hsu@HUHEPL.bitnet, Hsu@HSUNEXT.Harvard.edu}
   \vskip 0.3 cm
   {\it Lyman Laboratory of Physics,
	Harvard University,
	Cambridge, MA 02138}\\ }
  \vskip 0.3 cm
\end{center}

\vskip .5 in
\begin{abstract}
We examine recent claims that nonperturbative effects can prevent
the decoupling of a heavy fermion whose mass arises from a Yukawa
coupling to a scalar field. We show that in weakly coupled, four
dimensional models such as the standard model with heavy mirror
fermions the effects of the heavy  fermions can always be
accounted for by {\it local} operators involving light fields.
We contrast this with the case of the 1+1 dimensional Abelian
Higgs model, in which there does not appear to be a local effective field
theory
describing the low energy physics when heavy fermions are integrated out.

\end{abstract}
\end{titlepage}

\renewcommand{\thepage}{\arabic{page}}
\setcounter{page}{1}
\section{Decoupling and heavy fermions}
Perturbative considerations \cite{AC} lead us to believe that heavy particles
 can be decoupled from low energy degrees of freedom. We expect that the
 lagrangian describing the low energy degrees of freedom is affected by heavy
 particles only through renormalization effects and higher dimension operators
  which become negligible as the particle is made infinitely massive.
These considerations are central to the idea of effective field theory
\cite{EFT}
(for a
recent review, see \cite{HG}). Briefly stated, the idea of effective field
theory is that the S-matrix for light degrees of freedom can always be encoded
in terms of an effective lagrangian containing all possible renormalizable
plus higher dimension terms involving light fields, consistent with unbroken
symmetries\footnote{This point of view was emphasized by S. Weinberg in his
1993 Loeb Lectures.}. The mass scale suppressing the higher dimension
operators indicates where the effective theory breaks down, and usually
corresponds to  a heavy particle mass threshold.
In particular, nonlocal terms are not considered except insofar as the higher
dimension terms can be considered the expansion of some heavy particle
propagator.

It has been known for some time that there are some exceptions to the
expectations of perturbative decoupling.
In particular, if the heavy particle is a chiral fermion which participates in
anomaly cancellation, its decoupling can leave behind a Wess-Zumino term which
cancels the apparent anomalies of the remaining low energy degrees of freedom
\cite{DF,JP}. However, the Wess-Zumino term is a local functional of the light
fields and therefore does not violate the expectations of effective field
theory.
Recently, Banks and Dabholkar \cite{BD} have claimed to exhibit models in
which heavy particles do {\it not} decouple from low energy physics, and in
fact manifest themselves {\it nonlocally} in the low energy effective
lagrangian. The models in question exhibit topologically nontrivial field
configurations such as instantons. The otherwise heavy fermion fields have
zero modes in the background of these field configurations, and this is the
source of possible nondecoupling. While the models under consideration in
\cite{BD} are admittedly fine-tuned and unnatural, the violation of decoupling
is nonetheless interesting from a theoretical viewpoint as a test of the
effective field theory ideas discussed above.

In this letter we will reexamine the claims of Banks and Dabholkar. Two types
of models are discussed in \cite{BD}: the standard model plus mirror fermions
(motivated by considerations of lattice fermion doubling) and the Abelian
Higgs model in 1+1 dimensions. We find that it is only in the two dimensional
model that an actual violation of decoupling occurs. This is due to the
peculiar infinite range effects of instantons in this model, which restore
the otherwise broken gauge symmetry and alter the asymptotic states.
In the four dimensional
case the effects of the heavy fermions on the S-matrix for light states can be
reproduced by higher dimensional operators. However, we find that although the
S-matrices for light states can be made to agree in the full and effective
theories, the Green's functions or correlators, cannot.
Since it is the S-matrix that is measurable and physical, this distinction,
although
curious, is unimportant from the viewpoint of effective field theory.

\renewcommand{\thepage}{\arabic{page}}
\section{EFT for the vector standard model}

The investigations of Banks and Dabholkar were motivated by the problem of
fermion doubling in lattice versions of the standard model. The issue is
whether doublers can be decoupled from the low energy physics by giving them
masses of order the cutoff (e.g. via a Wilson-Yukawa type interaction)
\cite{B}.
In \cite{BD} the question is addressed in several continuum
models, the first of which is the standard model plus mirror fermions. By
mirror fermions we mean left handed Weyl fermions transforming in the complex
conjugate representations
of the standard model fermions. The resulting model
is thus vectorlike, but a large mass hierarchy between the standard model
fermions and the mirrors can be maintained if the standard model Yukawa
couplings are scaled down by a factor $f$ (say, $f \sim 10^{-2}$), with the
mirror Yukawa couplings kept $\sim 1$. In order to keep the gauge bosons and
Higgs particle in the low energy sector, we also rescale the renormalized gauge
and scalar
self couplings by the factor $f$. By performing the above rescalings we retain
the use of weak coupling perturbation theory at the cost of having to fine
tune enormously to keep the Higgs boson light in the face of large radiative
corrections from the mirror fermions.
(If the physical Higgs boson is not kept in the low energy sector,
there will be no topologically nontrivial configurations
with finite action in the low energy sector because such
configurations require a zero of the Higgs field \cite{finiteaction}.)
Note that the fermions that are decoupled in this model (and in the Abelian
Higgs
models studied in the next section) do not leave behind an uncanceled gauge
anomaly.
We therefore do not expect to find a Wess-Zumino term in the
effective theory, which remains completely gauge invariant.
Each of the terms appearing in the
effective Lagrangian must be gauge invariant, and hence the renormalizable
terms (dimension 4 or less) are completely specified as the usual ones, up to
possible renormalization of the couplings.

Now consider nonperturbative baryon number violation in this model.
In particular, consider the baryon violating decay of the (rescaled)
deuteron, originally studied by 'tHooft \cite{TH}.
In the low energy sector there appear to be instanton solutions
mediating such decays. On the
other hand, since the difference between ordinary and mirror baryon number is
exactly conserved, we know that in the full theory such a decay is
strictly forbidden. Both heavy and light fermions have zero modes in
the $SU(2)$ instanton background, and therefore the functional
determinant is zero for any topologically nontrivial process that
involves only light modes.
There appears to be a conflict between the exact and low energy
descriptions of the model!

Furthermore, we find other discrepancies between the full and
low energy theories when we consider correlators or Green's
functions. Imagine computing a correlator which contributes
to some baryon number conserving process in the low energy
sector. In general there will be contributions to such processes
from instanton-antiinstanton configurations. Because of the zero
modes mentioned above, each (anti)instanton vertex must (destroy)
create a fermion of each flavor.
Since in the exact theory there are no perturbative
interactions which violate baryon number, each fermion
line must connect the instanton and antiinstanton.
Thus, configurations in which the pair are widely separated
are suppressed by
exponential factors $exp(- M_F |x-y|)$, where $M_F$ is
the mass of the heavy
fermions and $|x-y|$ is the separation. These factors arise
from the coordinate space heavy
fermion propagators which connect the instanton and antiinstanton.
 Now consider the low energy theory. Since there are no heavy
fermion fields (and therefore no heavy fermion zero modes) in the effective
theory, this ``confinement'' of instantons (or topological charge in general)
is absent. Correlators in the zero
topological charge sector will receive contributions from
instanton-antiinstanton pairs which are widely separated compared to the heavy
fermion Compton wavelength $M_F^{-1}$. If the only additional terms induced in
the low energy theory by integrating out heavy fermions are local functions of
the light fields, suppressed by powers of $1/M_F$, it is hard to see how these
additional terms can lead to confinement of instantons. In particular,
the instanton field configurations are smooth functions,
and in the case of large instantons ($\rho_I >> M_F^{-1}$), the
effect of the higher dimension operators on the action of the instanton
can be considered a small perturbation.

Banks and Dabholkar claim that the resolution of these puzzles
is that the heavy fermions do not truly decouple from the low
energy physics (they become ``phantom fields'')
and that a correct description of low energy
physics requires some nonlocality of the effective theory.
They further speculate that this is a generic phenomena
associated with decoupling of heavy particles whose mass
arises from spontaneous symmetry breaking. Because the
heavy particles can become massless (or at least light)
in the background of a field configuration with vanishing
scalar vacuum expectation value, the logic is that
they do not decouple from low energy processes mediated by such configurations.

Here we argue that the opposite is true:
the effects of the virtual heavy fermions can be reproduced by
appropriately chosen local operators of the
light fields. More specifically, the S-matrices of the full and effective
theory can be matched by appropriate choice of these operators. While this
resolves issues of the first type mentioned above, e.g. deuteron decay, we
will see that it does not resolve discrepancies in correlators or Green's
functions between the full and effective theories. This is acceptable from the
standpoint of quantum field theory, as the S-matrix is the fundamental object
that is related to experimental measurement. On the other hand, from the
statistical mechanical viewpoint it is the entire correlator that may be of
interest, not just its
asymptotic behavior which determines the S-matrix in field theory.

It is straightforward to match S-matrix elements between the
exact and effective theories. The idea is that given any S-matrix
element in the exact theory, involving only light in- and out-states,
one can choose higher dimension operators in the effective theory
to achieve matching. In the vector standard model the S-matrix elements
in question are those that violate baryon number without violating
mirror baryon number. In the exact theory, these S-matrix elements
are exactly zero. In the low energy theory there are contributions
from two sources: instantons and explicit baryon number violating operators
such as
\Eq
\O_{12} = \prod_{I=1}^{N_F} \epsilon_{ij} \epsilon_{kl} \epsilon_{\alpha \beta
\gamma}
(q^{\alpha}_i q^{\beta}_j q^{\gamma}_k l_l )_I,
\Endl{Op}
where $\{i,j,k,l\}$ are SU(2) indices, $\{\alpha, \beta, \gamma \}$
are color indices and $I$ is a flavor index.
The instantons\footnote{By instantons here we really mean constrained
instantons
\cite{TH,Affleck} since there are no exact instantons in a spontaneously
broken theory. The constrained instantons exist at all sizes $\rho$,
with those of size $\rho \sim v^{-1}$ giving the dominant contribution to the
path integral.}
also contribute to correlators such as
\Eq
\langle \prod_{I=1}^{N_F} ( q(w_I) q(x_I) q(y_I) l(z_I))_I \rangle
\Endl{corr}
which yield S-matrix elements when the LSZ (Lehmann-Symanzik-Zimmermann)
projection \cite{text} is applied:
\Eq
\langle i |S| f \rangle = LSZ \{\langle \prod_{I=1}^{N_F} ( q(w_I) q(x_I)
q(y_I) l(z_I))_I \rangle \}
\Endl{LSZ}
Recall that the LSZ projection removes all but the coefficient of the
multiple on-shell pole of the correlator. In momentum space,
\Eq
LSZ =  C \prod_{i=1}^{N_{f}} (i \PS_i + m ) \prod_{j=1}^{N_{b}} ( p^2_j + M^2)
\End
with appropriate combinatorial factor $C$.
In order to reproduce the exact result of zero (i.e. a stable deuteron)
the operator $\O_{12}$, and similar ones involving additional derivatives,
must be adjusted to cancel the instanton contribution to (\ref{LSZ}).

Let us study this more explicitly.
Let
\Eq
G_E(x_1,...,x_N) = \langle \prod_{I=1}^{N_F}
( q(x_1) q(x_2) q(x_3) l(x_4))_I \rangle_{Exact}
\Endl{exact}
be evaluated in the full theory, while
\Eq
G_{NP} (x_1,...,x_N) = \langle \prod_{I=1}^{N_F}
( q(x_1) q(x_2) q(x_3) l(x_4)_I \rangle_{Nonperturbative}
\Endl{ren}
is the contribution from nonperturbative field configurations computed
in the effective theory. In the effective theory, there will also be a
direct perturbative contribution
$G_{\O} (x_1,...,x_N)$
from higher dimension operators $\O_n$.
Now consider matching the contributions to the light
S-matrix from $G_E$ and from $G_{NP}$ plus $G_{\O}$ by appropriate choice of
operators $\O_n$
containing N light fields and any number of derivatives.
We want
\Eq
\langle i |S_{E}| f \rangle
= \langle i |S_{NP}| f \rangle + \langle i |S_{\O}| f \rangle,
\End
where
\Eqa
\langle i |S_{E,NP,\O}| f \rangle
&=& \langle k_1,..,k_m | S_{E,NP,\O} | k_{m+1},..,k_N \rangle \\
&=&  LSZ~[ \prod_{i=1}^N
\int~ d^4x_i~ e^{ik_i \cdot x_i}~ ( G_{E,NP,\O}( x_1,..,x_N ) )     ]\\
   &=& f_{E,NP,\O} (k_1,..,k_N).
\Enda
Here $f_{E,NP,\O} (k_1,..,k_N)$ are simply the coefficients
of the multiple poles in the Fourier transforms of
$G_{E,NP,\O}$. For Euclidean momenta with imaginary parts less
than the light particle masses these residues are completely analytic,
so a power series representation is always possible. We can therefore expand
each
$f_{E,NP,\O} (k_1,..,k_N)$ in the variables $k_i^2, k_i \cdot k_j$.

In the vector standard model $f_E = 0$ while $f_{NP}$ is determined by the
currently unknown behavior of anomalous baryon number violating
amplitudes at arbitrary momenta $k_i$ \cite{RWE}.
As for $f_{\O}$, each operator $\O_n$ contributes a term to
$f_{\O}$ with a power of $k_i$ for each derivative
$\partial_i$ in the operator. The coefficients of the
operators are then determined by equating the terms in
the power series expansions of $f_{NP}$ and $f_{\O}$.

It is important to emphasize that this matching must be order
by order in some small parameter. This is because the
nonperturbative correlators in the effective theory depend
on the coefficients of the operators $\O_n$. However, these
effects will be higher order in the small parameter. In a
weakly coupled theory like the vector standard model, the small parameter is
$e^{-S_0}$ (the instanton action) times powers of the
ratio of light and heavy scales. In a strongly coupled
model the exponential factor may be absent but the second factor will remain.

It should be clear that the construction considered above
is quite general. As long as we know the asymptotic (in-, out-)
states of the low energy sector, we can reproduce the
exact S-matrix between light fields by suitable choice
of higher dimension operators $\O$. The point is that
any S-matrix element between N light fields will get a
direct contribution from the corresponding operator $\O_N$.
The momentum dependence of the S-matrix is determined by
operators with derivatives acting on $\O_N$. In the case
at hand, the higher derivative operators are chosen to
reproduce the ``form factors'' of instanton amplitudes.

Now we turn to the matching of correlators between the
exact and effective theories. In perturbative calculations
\cite{HG} one often defines the effective theory by
requiring that Green's functions match. Technically,
this is of course too stringent a requirement. We are
familiar with the result that in ordinary gauge theory
different choices of gauge lead to different Green's
functions that typically disagree, while the corresponding
S-matrix elements agree. Clearly, Green's functions are
not by themselves physical objects. Here we will argue
that, at least as far as the nonperturbative effects
in the vector standard model are concerned, it is impossible
to match Green's functions.

Suppose that we were able to impose that all correlators of
a finite number of fields match exactly between the two
theories. Consider the following correlator, that of a functional delta
function:
\Eqa
\langle \delta [ \phi(x) - \phi_0 (x) ] \rangle
&=& N \int  D\phi~ \delta[ \phi(x) - \phi_0 (x) ] ~ e^{-S[\phi]} \\
&=& N~e^{-S[\phi_0 (x) ] },
\Endla{delt}
where here we use $\phi(x)$ generically to represent
all fields. Now, we can represent
$ \delta[ \phi(x) - \phi_0 (x) ] $ in terms of a
finite number of standard delta functions if we discretize spacetime:
\Eq
\delta[ \phi(x) - \phi_0 (x) ] = \prod_x \delta( \phi(x) - \phi_0 (x) ).
\End
Furthermore, the standard delta functions can be
expanded in a power series in
$( \phi(x) - \phi_0 (x) )$. The sequence
$\delta_n ( \phi(x) - \phi_0 (x) )
= \frac{n}{\sqrt{\pi}} e^{- n^2 ( \phi(x) - \phi_0 (x) )^2}$
converges to a delta function, and since $e^{-z^2}$ is analytic
everywhere except $z = \infty$ it can be expanded in a uniformly
convergent series in $z^2$. Therefore, to any desired accuracy
the expectation of the delta functional (\ref{delt}) can be represented in
terms
of a finite number of correlators of a finite number of fields.
If these correlators agree, then (up to a field
independent constant) the actions of any chosen field
configuration $\phi_0 (x)$ must agree in the exact and effective theories.

As mentioned previously, in the exact theory there is a
long range ``interaction'' (if we think in terms of
statistical mechanics, with energy substituted for
action) between widely separated lumps with nontrivial
topological charge. The interaction grows with
separation $|x-y|$ and is therefore confining.
Given the arguments of the previous paragraph
it is clear that correlator matching requires
that this effect be reproduced in the effective theory.

It seems clearly impossible to induce a confining, long
range interaction between lumps via local operators without
introducing new degrees of freedom. But, for the sake
of clarity, let us further belabor the point.
Consider the effective lagrangian constructed so as to match S-matrix elements
\Eq
{\cal L}_{eff} = {\cal L}_{ren} + {\cal L}_{pert} + {\cal L}_{NP}.
\End
Here the term ${\cal L}_{ren}$ consists of renormalizable
terms while ${\cal L}_{pert}, {\cal L}_{NP}$ contain
the perturbative and nonperturbative higher dimensional
operators required for matching. In particular, all the
terms in ${\cal L}_{pert}$ conserve (B+L) number, while
all the terms in ${\cal L}_{NP}$ violate it. Both types
of terms are suppressed by powers of the ratio of light
to heavy scales, and the nonperturbative terms also carry a suppression of
$e^{-S_0}$.

Now consider a widely separated pair of lumps at coordinates
$x$ and $y$. Let $B_x$ and $B_y$ be balls centered at $x$
and $y$ with radii much less than $|x-y|$. In order to
reproduce the long range interaction we require that
\Eq
\Delta S = \int_{R^4 - B_x - B_y} ~~d^4x~ \{ [{\cal L}_{pert}
+ {\cal L}_{NP}]_{lumps} -  [{\cal L}_{pert}
+ {\cal L}_{NP}]_{vacuum} \}~ \sim~ M_F |x-y|
\Endl{lump}
However, we can arrange that the fields approach their
vacuum values to exponential accuracy far from the
centers of the lumps. Since the field configurations
are smooth, and the terms in ${\cal L}_{pert} + {\cal L}_{NP}$
are by assumption simply polynomials in the fields,
we can arrange for the left hand side of (\ref{lump})
to be arbitrarily small for large separation.
We are thus clearly {\it unable} to reproduce the linear potential.

\renewcommand{\thepage}{\arabic{page}}
\section{EFT in two dimensions}

Banks and Dabholkar also consider nondecoupling in the
Abelian Higgs model in two dimensions. In two dimensions
and at large-N (where N is the number of fermion flavors)
they were able to explicitly construct a model with an
arbitrarily large hierarchy between the heavy fermion
mass and the scale of the light degrees of freedom
(scalars, gauge bosons and light fermions). Again in
this model there are nonperturbative effects which are
 suppressed by the presence of the heavy fermions,
 regardless of their mass. Here we argue that there
 is indeed a breakdown of effective field theory in the
 two dimensional case. In contrast with the vector
 standard model, instantons in the Abelian Higgs (AH)
 model have infinite range effects which cannot be
 reproduced even in the S-matrix by the introduction of local operators.

Consider the following lagrangian,
\Eqa
{\cal L} &=& \frac{-1}{4e^2} F_{\mu \nu}^2 +
|D_{\mu} \phi|^2 - V(\phi^* \phi) \nonumber \\
   &+& \bar{\psi} \{ i \gamma^{\mu}
   [ \partial_{\mu} - q [\frac{1 - \epsilon \gamma_3}{2} ] A_{\mu} ]
   + g(\phi^*)^q[ \frac{1 - \epsilon \gamma_3}{2} ] +
   g(\phi)^q[ \frac{1 + \epsilon \gamma_3}{2} ]
       \} \psi
\Enda
where $q$ and $\epsilon$ are $3 \times 3$ matrices:
$q = diag(q_1,q_2,q)$ and $\epsilon = diag(1,1,-1)$.
The charges are chosen to satisfy the anomaly cancellation
condition $q^2 = q_1^2 + q_2^2$. Note that there are
two left moving and one right moving fermion of charges
$q_1, q_2, q$ which are paired with singlets by the Yukawa couplings.
It is important for our purposes that at least some of the fermions carry
fractional charges, so that they cannot be screened by the scalar, which
carries unit charge.
As noted in \cite{BD}, the fermions $\psi$ exhibit
zero modes in topologically nontrivial backgrounds.
Therefore, instanton amplitudes must be accompanied
by the creation or destruction of fermion number.
The large-N limit is obtained by including N copies
 of the heavy fermion field and rescaling the couplings
 so that $e^2 N = E^2, g^2 N = G^2$ and
 $\lambda N = \kappa$ remain fixed as $N \rightarrow \infty$.
The Euclidean AH lagrangian is identical to the
Landau-Ginzburg Hamiltonian for a two dimensional
superconductor. In this context the instantons are simply
vortices and the dilute instanton gas represents the statistical mechanics of
these vortices.

It is again clear from arguments similar to those of the previous
section that effects like the confinement of instantons cannot be
reproduced by a local effective lagrangian. However, in this model
confining instantons does more than just prevent matching of correlators.
In the AH model the instantons are crucial to determining the
realization of the gauge symmetry as well as the asymptotic states
of the model. In the absence of fermion zero modes, it is easy to
see that in a sufficiently large box the effect of instantons is
to restore the otherwise spontaneously broken $U(1)$ symmetry.
Consider the vacuum expectation value
(for simplicity we assume that the vacuum angle $\theta = 0$)
\Eq
\langle  \phi(x) \rangle = N \int ~D\phi~ \phi(x) e^{-S[\phi]}
\End
In perturbation theory, we would find that
$\langle \phi(x) \rangle \neq 0$, therefore signalling spontaneous
symmetry breaking. In two dimensions, however, instantons
can restore the symmetry. A simple way to see this is to
consider the analogous Landau-Ginzburg model. In the
Landau-Ginzburg model this correlator is equivalent
to the average value of $\phi(x)$ in the presence of
vortices at low temperature. A vortex centered at $y$ very
far from $x$ exerts an effect on $\phi(x) = v e^{i \theta_{xy}}$,
regardless of the separation $| x-y |$. Here $\theta_{xy}$ is
the angle between the line $\overline{xy}$ and an arbitrary
line emanating from $y$. An anti-vortex has a similar effect
except with opposite angle. Now, although vortices are extremely
rare at low temperature, occurring with exponentially small density,
in an infinite volume there are nevertheless an infinite number of
vortices and anti-vortices. Each of these, centered at arbitrary
points $y_i$, completely disorder the expectation value
$\langle \phi(x) \rangle$ yielding zero in the infinite volume limit.
In the dilute vortex approximation, we find
\Eq
\langle  \phi(x) \rangle = v~ exp[ - k \int d^2R~ (2 - \frac{\phi^+ (x-R)}{v}
    - \frac{\phi^- (x-R)}{v} ) ],
\Endl{vortex}
where $k \sim v^2 e^{-S_0}$ is the density of vortices, and $\phi_{+,-}$ is the
vortex or antivortex solution. Since $\int d^2R~ [\phi_{+,-} (x-R)] = 0$,
(\ref{vortex})
goes to zero in the infinite volume limit.

This restoration of symmetry at sufficiently large distances suggests
that a Coulomb interaction is present at large distances between charged
particles. But in one spatial dimension a Coulomb interaction is confining,
and therefore leads to the requirement that asymptotic states in the
AH model with instantons be {\it neutral}, rather than charged, as
would be expected from the perturbative Higgs mechanism. Indeed, one
can verify that the Wilson loop for fractional charge $q$
\Eq
\langle W \rangle ~ =~  \langle e^{ q \int dx^{\mu} A_{\mu} } \rangle
{}~=~  \langle e^{ q \int d^2x~ \epsilon_{\mu \nu} F^{\mu \nu}} \rangle
\End
displays area law behavior in the presence of instantons \cite{Sid},
thus signalling charge confinement.

The point of the previous comments is that the instantons in this
model actually alter the asymptotic states of the model. When heavy
fermions are introduced which exhibit zero modes in the instanton
background, they suppress the instantons and therefore also prevent
the restoration of the $U(1)$ symmetry. There are clearly no local
operators which can mimic the effect of the heavy fermions:
they would either have to induce a confining interaction between
instantons (already ruled out by previous arguments),
or eliminate topologically nontrivial configurations
entirely from the low energy sector.

The existence of nontrivial configurations such as instantons in the
low energy theory can be guaranteed on topological grounds as long
as the  Euclidean effective lagrangian remains positive definite.
In that case, we continue to require that terms in the action like
$|D_{\mu} \phi|^2$ and $F_{\mu \nu}^2$ approach zero at infinity.
This condition is sufficient to guarantee the topological
classification of configurations that leads to instantons or
vortices. It is easy to see that the effect of {\it exactly}
integrating out the heavy fermions leads to corrections to
the effective action which leave it positive definite.
The heavy fermion contribution to the exact effective action is\footnote{Note
that
this object requires careful definition \cite{reg}.
It does not contain a Wess-Zumino term because the
gauge anomalies of the $\psi$'s have been chosen to cancel.}
\Eqa
\Gamma_{\psi} [\phi] &=& - ln ( \int D\psi~ e^{-S[\phi,\psi]} ), \\
      &=& - ln ( \frac{1}{C} \int D\psi~ e^{-S[\phi,\psi]} ) - ln(C)
\Endla{ea}
where $\phi$ is meant to represent all of the light fields,
and $C$ is a $\phi$ independent constant. $C$ is chosen by
regulating the path integral in (\ref{ea}), and requiring
\Eq
||~ \int D\psi~ e^{-S[\phi,\psi]} ~ || < C,~~ \forall \phi.
\End
With this choice of $C$, $\Gamma_{\psi}[\phi]$ is positive
definite up to a $\phi$ independent constant.
If a convergent expansion of the effective action in local
operators exists, it must also be positive definite. Hence
 we expect the effective theory to contain instantons.

It does not seem possible that the interactions induced by
heavy fermions render the action of all topologically
nontrivial configurations infinite, since for smooth
configurations the effect of the higher dimension operators
is suppressed by powers of the ratio of the light to heavy
scale, which can be made arbitrarily small at large N.
For any finite value of the vortex energy (or instanton action),
however large, our previous arguments for symmetry
restoration at infinite volume remain valid. Thus the
spectrum of the low energy theory will always consist
only of charge singlet states, as opposed to the spectrum
of the exact theory with massive fermions, which
exhibits the Higgs mechanism.
Therefore in the AH model we have a bona fide
counterexample to the idea of effective field theory:
no local effective lagrangian can reproduce the exact
S-matrix of this model in the light sector.

\renewcommand{\thepage}{\arabic{page}}
\section{Final Comments}

We have seen that, at least as far as nonperturbative effects are concerned,
there is an important distinction to be made between matching
the Green's functions and matching the S-matrices of exact
and effective low energy theories. We find that in the
vector standard model studied by Banks and Dabholkar, the
latter is possible while the former is not. Because the exact S-matrix
can be reproduced, this model
does not present a breakdown in effective field theory.

However, we do find, in agreement with Banks and Dabholkar, that in the
two dimensional Abelian Higgs model it is impossible to reproduce
the exact S-matrix in the low energy sector via a local effective lagrangian.
This seems to be a result of the long range effects of topological defects in
this
class of models, and is clearly not a generic phenomena. It would appear that
the
necessity of ``phantom fields'' is limited to models of this sort, and does not
extend to all models in which fermions receive their masses from Yukawa
couplings.

We comment briefly on the relevance
of these conclusions to the original lattice motivations
for \cite{B,BD}.
The point of Banks' original arguments was that attempting to decouple fermion
doublers
in gauged chiral models by giving them masses of order the cutoff was doomed to
failure insofar
as the resulting model would not exhibit low energy baryon number violation.
This
conclusion remains valid, but it is instructive to note that from the
low energy viewpoint the absence of baryon number violation is due to the
presence
of certain higher dimension operators which explicitly violate baryon number
and which
cancel the low energy nonperturbative effects exactly.
In other words the low energy limit of the model with heavy doublers is
essentially the
conventional standard model plus some very small higher dimension operators.
If the purpose of the hypothetical
lattice simulation is to determine nonperturbative quantities such as the size
and
energy dependence of baryon number violation, then it is true that the
Wilson-Yukawa
method is insufficient\footnote{For a discussion of other problems inherent in
Wilson-Yukawa
models, see \cite{lattice}.}.
On the other hand, other dynamical properties of the system (such
as baryon number {\it conserving} correlators) which may be relevant to issues
such as
the phase structure of the chiral model, will only suffer small
corrections due to the induced operators.

Finally, we would like to mention another point of view advanced by Georgi,
Kaplan and Morrin (GKM) \cite{GKM}.
These authors argue that the full and effective theories can be reconciled if
it can be shown that there are
{\it no} instantons in the low energy theory - in other words that some of the
induced
higher dimension operators render the instanton action infinite. This
possiblity was mentioned above, but
discounted.  Since even a `large' instanton must have a zero of the Higgs field
(due to topological considerations), GKM argue that it is possible that some of
the induced
higher dimension operators are singular when evaluated on such a zero. This
possibility
in itself seems only to
render the action of topologically nontrivial configurations incalculable in
the low energy theory, not
necessarily infinite. It is further disturbing that in such a picture the low
energy physics is sensitive to the
behavior of the fields at arbitrarily short distances (i.e. - near the Higgs
zero). However, if this picture
is correct, it would imply that mirror fermions can be accounted for in any
number of dimensions without
nonlocality, or even anomalous higher dimension operators.

\vskip .5 in

The author would like to acknowledge Steven Weinberg, whose 1993 Loeb Lectures
on
Effective Field Theory stimulated this investigation.
He would also like to thank E. Farhi, H. Georgi, S. Hughes, L. Kaplan, D.
Morrin and
S. Osofsky for useful discussions.
SDH acknowledges support from the National Science Foundation under grant
NSF-PHY-92-18167,
the state of Texas under grant TNRLC-RGFY93-278B,
and from the Harvard Society of Fellows.

\end{document}